\documentclass[aps,pra,twocolumn,groupedaddress,showpacs]{revtex4-1}
\usepackage{graphicx}
\usepackage{amsmath}
\usepackage{textcomp}
\usepackage{mathrsfs}
\usepackage{amsfonts}

\begin{document}

\newcommand{\vett}[1]{\textbf{#1}}
\newcommand{\uvett}[1]{\hat{\textbf{#1}}}
\newcommand{\fieldE}[2]{E(\vett{#1},#2)}
\newcommand{\fieldA}[2]{A(\vett{#1},#2)}

\newcommand{\fieldX}[3]{\psi_{#1,#2}^{(#3)}(R,\zeta)}
\newcommand{\fieldXc}[3]{\psi_{#1,#2}^{(#3)*}(R,\zeta)}

\newcommand{\fieldXz}[3]{\psi_{#1,#2}^{(#3)}(R,z)}
\newcommand{\fieldXcz}[3]{\psi_{#1,#2}^{(#3)*}(R,z)}

\newcommand{\fieldXo}[2]{\psi_{#1,#2}^{(0)}(r,X)}
\newcommand{\fieldXco}[2]{\psi_{#1,#2}^{(0)*}(r,X)}

\newcommand{\fieldXp}[3]{\psi_{#1,#2}^{(#3)}(R',\zeta)}

\newcommand{\fieldXa}[3]{\psi_{#1,#2}^{(#3)}(R,\zeta_1)}
\newcommand{\fieldXb}[3]{\psi_{#1,#2}^{(#3)}(R,\zeta_2)}

\newcommand{\creatX}[3]{\hat{a}_{#1,#2}^{\dagger}(#3)}
\newcommand{\annX}[3]{\hat{a}_{#1,#2}(#3)}
\newcommand{\creatXb}[3]{\hat{b}_{#1,#2}^{\dagger}(#3)}
\newcommand{\annXb}[3]{\hat{b}_{#1,#2}(#3)}
\newcommand{\annA}[2]{\hat{A}(\vett{#1},#2)}
\newcommand{\creatA}[2]{\hat{A}^{\dagger}(\vett{#1},#2)}

\newcommand{\beq}{\begin{equation}}
\newcommand{\eeq}{\end{equation}}
\newcommand{\barr}{\begin{eqnarray}}
\newcommand{\earr}{\end{eqnarray}}
\newcommand{\bseq}{\begin{subequations}}
\newcommand{\eseq}{\end{subequations}}
\newcommand{\bal}{\begin{align}}
\newcommand{\eal}{\end{align}}
\newcommand{\ket}[1]{|#1\rangle}
\newcommand{\bra}[1]{\langle #1|}
\newcommand{\expectation}[3]{\langle #1|#2|#3 \rangle}
\newcommand{\braket}[2]{\langle #1|#2\rangle}

\newcommand{\multi}{\{\vett{m}\}}
\newcommand{\multid}{\{\vett{a}\}}
\newcommand{\multit}{\{\vett{b}\}}

\newcommand{\annF}[3]{\hat{\phi}_{#1#2}(#3)}
\newcommand{\creatF}[3]{\hat{\phi}_{#1#2}^{\dagger}(#3)}
\newcommand{\annFd}[2]{\hat{\phi}_{#1#2}}
\newcommand{\creatFd}[2]{\hat{\phi}_{#1#2}^{\dagger}}

\newcommand{\creatXt}[3]{\hat{a}_{#1,#2}^{\dagger}(#3,t)}
\newcommand{\annXt}[3]{\hat{a}_{#1,#2}(#3,t)}
\newcommand{\annFt}[3]{\hat{\phi}_{#1#2}(#3,t)}
\newcommand{\creatFt}[3]{\hat{\phi}_{#1#2}^{\dagger}(#3,t)}

\title{Quantum X Waves with Orbital Angular Momentum in Nonlinear Dispersive Media}

\author{Marco Ornigotti$^1$}
\email{marco.ornigotti@uni-rostock.de}
\author{Claudio Conti$^{2,3}$}
\author{Alexander Szameit$^1$}

\affiliation{$^1$Institut f\"ur Physik, Universit\"at Rostock, Albert-Einstein-Stra\ss e 23, 18059 Rostock, Germany}
\affiliation{$^2$University of Rome La Sapienza, Department of Physics, Piazzale Aldo Moro 5 00185, Rome, Italy}
\affiliation{$^3$Institute for Complex Systems, National Research Council, (ISC-CNR), Via dei Taurini 19, 00185, Rome (IT) }

\begin{abstract}
We present a complete and consistent quantum theory of generalised X waves with orbital angular momentum (OAM) in dispersive media. We show that the resulting quantised light pulses are affected by neither dispersion nor diffraction and are therefore resilient against external perturbations. The nonlinear interaction of quantised X waves in quadratic and Kerr nonlinear media is also presented and studied in detail.
\end{abstract}

\pacs{00.00, 20.00, 42.10}
\maketitle

\section{Introduction}

Electromagnetic waves are usually subject to diffraction and dispersion, i.e., a progressive broadening during propagation of the wave in both space and time, respectively. Ultimately, these effects are connected with the bounded nature of the wave spectrum and, therefore, to its finite energy content \cite{jackson}. Maxwell's equations, however, admit diffraction- and dispersion-free solutions, the so-called localised waves \cite{localizedWaves}. An example of such solutions in the monochromatic domain are the well known Bessel beams \cite{durnin}. In the pulsed domain, the most famous representatives of localised waves are X waves. Firstly introduced in acoustics by Lu and Greenleaf in 1992 \cite{ref2,ref3}, they have been the subject of an extensive study in different areas of physics, such as nonlinear \cite{ref4,ref5} and quantum \cite{ref7} optics, condensed matter physics \cite{ref6}, integrated optics \cite{ref8,ref9} and optical communications \cite{ref10}, to name a few. A comprehensive review of the topic can be found in Refs. \cite{localizedWaves} and \cite{nondiffractingWaves}. 

Traditionally, X waves are understood as superpositions of Bessel beams of zero order, therefore neglecting their possible orbital angular momentum (OAM) content. The latter, in fact, is know to be related to the twisted phase front of higher order Bessel beams \cite{OAMbook}. Generalisation of the traditional X waves to the case of OAM-carrying X waves has been only recently investigated \cite{ref11}. 

Despite the great amount of work that has been done in the subject in the last decades, however, investigations of quantum properties of X waves are very few and limited to the case of no OAM content \cite{ref7,claudio1,claudio2}. Very recently, the effect of the OAM content of X waves in squeezing processes has been analysed in detail, highlighting new features and possibilities \cite{ref12}. However, a comprehensive quantum theory of X wave and a complete analysis of their properties and dynamics at the quantum level has only be sketched in Ref. \cite{ref12} and not fully developed. 

In this work, therefore, we present a complete and comprehensive quantum theory of X waves with OAM. Although the quantisation of the electromagnetic field carrying angular momentum has already been carried out in both the context of quantum field theory \cite{quant1} and for generalised Gaussian-Airy wave packets \cite{quant2}, the approach presented here constitutes a more general framework, where dispersion effects and nonlinearities are automatically accounted for. In particular, we discuss in detail the dynamics of quantum X waves in media exhibiting  $\chi^{(2)}-$ and $\chi^{(3)}-$nonlinearities. Our findings reveal that photon pairs generated via parametric down conversion present continuous variable entanglement in their velocity (i.e., Bessel cone angle) degree of freedom, while their OAM content plays an active role in the determination of the properties of photons propagating in Kerr media. In the latter case, the Kerr nonlinearity introduces a coupling between different OAM states, whose strength is essentially regulated by the overlap integral between the different X waves modes involved in the process. The approach presented here is limited to the quantisation of a scalar pulse propagating in a nonlinear medium. A complete discussion of the effect of vector properties of X waves is therefore left to future works. 

This work is organised as follows: in Sect. II we derive the expression of the wave equation in dispersive media casted as as a time evolution problem and we present its general solution in terms of X waves. Section III is instead dedicated to the quantisation of such X waves and to the study of their main properties. In this section, moreover, particle states and coherent states of OAM-carrying X waves are defined and it is shown that X wave coherent states carry finite energy. In Sect. IV, we consider $\chi^{(2)}$ nonlinearities involving quantum X waves carrying OAM, and we discuss in particular the case of squeezing and entanglement. Section V is instead devoted to the study of quantum X waves with OAM in the presence of a Kerr nonlinearity. Finally, conclusions are drawn in Sect. VI.

\section{Wave Equation in Dispersive Media}
As a starting point of our analysis, le us consider the propagation of a scalar electromagnetic field in a linear, dispersive medium, characterised by the refractive index $n=n(\omega)$
\beq\label{eq1}
\left(\nabla^2-\frac{n^2}{c^2}\frac{\partial^2}{\partial t^2}\right)\fieldE{r}{t}=0.
\eeq
Moreover, let us assume that the paraxial approximation applies and that we can write the electric field as
\beq\label{eq2}
\fieldE{r}{t}=\left(\frac{\varepsilon_0n^2}{2}\right)^{-1/2}\fieldA{r}{t}e^{i(kz-\omega t)},
\eeq
where $k=n\omega/c$ and the normalisation constant has been chosen such that the total energy of the field (i.e., the field intensity) can be written as
\beq\label{eq3}
\mathcal{E}=\frac{\varepsilon_0n^2}{2}\int\,d^3r\, |\fieldE{r}{t}|^2=\int\,d^3r\,|\fieldA{r}{t}|^2.
\eeq
Direct substitution of this Ansatz into Eq. \eqref{eq1} results in the conventional Fock-Leontovich equation, namely \cite{nondiffractingWaves}
\beq\label{eq3bis}
2ik\frac{\partial A}{\partial z}-2i\frac{n^2\omega}{c^2}\frac{\partial A}{\partial t}+\nabla_{\perp}^2A-\frac{n^2}{c^2}\frac{\partial^2A}{\partial t^2}=0.
\eeq
The above equation describes a propagating wave along the $z$-direction. However, this problem can be equivalently casted in terms of a time evolution problem, thus allowing a generalisation of the results of this investigation to all problems that admit evolution equations that can be casted in terms of Schr\"dinger-like equations. To do that, we introduce the reference frame $\{\zeta=z-(c/n)t,\tau=t\}$, co-moving with the envelope itself, and rewrite Eq. \eqref{eq3bis} as follows:
\beq\label{eq5}
2i\omega\frac{\partial A}{\partial\tau}+\nabla_{\perp}^2A-\frac{\partial^2A}{\partial\zeta^2}=\frac{n^2}{c^2}\frac{\partial^2A}{\partial\tau^2}-\frac{2n}{c}\frac{\partial^2A}{\partial\tau\partial\zeta}.
\eeq
Then, we apply the slowly varying envelope approximation (SVEA) to the envelope function $\fieldA{r}{\zeta}$, so that the right-hand side of the above equation can be neglected \cite{boyd}. If we now transform back into the reference frame $\{x,y,z,t\}$ we obtain the following equation
\beq\label{eq6}
i\frac{\partial A}{\partial t}+i\omega'\frac{\partial A}{\partial z}-\frac{\omega''}{2}\frac{\partial^2A}{\partial z^2}+\frac{\omega'}{2k}\nabla_{\perp}^2A=0,
\eeq
which is now describing the evolution in time of the field envelope $\fieldA{r}{t}$. Moreover, $\omega'$ and $\omega''$ are the first and second order dispersion defined as follows:
\bseq
\begin{align}
\omega'&=\frac{d\omega}{dk},\\
\omega''&=\frac{d^2\omega}{dk^2},
\end{align}
\eseq
where $k=k_0n(\omega)$. Equation \eqref{eq6} constitutes then the starting point of our investigation and will be used in the remaining of this manuscript to study the propagation of an electromagnetic field in a dispersive medium characterised by a refractive index $n(\omega)$ and first and second order dispersion $\omega'$ and $\omega''$, respectively. 

\subsection{Generalised X Wave Expansion}
We now look for solutions of Eq. \eqref{eq6} as superposition of X waves. To do that, let us first write the envelope function $\fieldA{r}{t}$ in terms of its 3D Fourier transform, namely
\beq\label{eq7}
\fieldA{r}{t}=\int\,d^3k\,\tilde{A}(\vett{k})\,e^{i\vett{k}\cdot\vett{r}}\,e^{-i\widetilde{\Omega}(\vett{k})t},
\eeq
where $\widetilde\Omega(\vett{k})$ reflects the fact that the frequency of the field envelope experiences dispersion while the field propagates in the medium. If we now introduce the cylindrical coordinates $\{R,\theta,z\}$ and $\{k_{\perp},\varphi,k_z\}$, the previous integral becomes
\beq\label{eq8}
\fieldA{r}{t}=\int\,dk_z\,\int\,d^2k\,\tilde{A}(\vett{k})\,e^{ik_{\perp}R\cos(\varphi-\theta)}\,e^{-i\widetilde{\Omega}(\vett{k})t}\,e^{ik_zz},
\eeq
where $d^2k=k_{\perp}dk_{\perp}d\varphi$. Following Ref. \cite{localizedWaves}, we choose the following explocit form for the Fourier spectrum $\tilde{A}(\vett{k})$ to represent the solution $A(\vett{r},t)$ as a superposition of X waves
\beq\label{eq9}
\tilde{A}(\vett{k})=\sum_{m=-\infty}^{\infty}d_mS(k_{\perp},k_z)e^{im\varphi},
\eeq
where $S(k_{\perp},k_z)$ is an arbitrary spectrum whose explicit form will be given below. Moreover, we also assume that 
$\partial\widetilde{\Omega}(\vett{k})/\partial\varphi=0$. Substituting the above spectrum into Eq. \eqref{eq8} and using the integral definition of thee Bessel function \cite{nist}, after a straightforward manipulation we arrive at the following form for the field envelope:
\barr\label{eq11}
\fieldA{r}{t}&=&\sum_md_m\int dk_ze^{ik_zz}\int_0^{\infty}dk_{\perp} k_{\perp}\nonumber\\
&\times&  S(k_{\perp},k_z)\, J_m(k_{\perp} R)\,e^{i[m\theta-\widetilde{\Omega}(\vett{k})t]}.
\earr
The above result constitutes an exact solution of Eq. \eqref{eq6} if $\tilde\Omega(\vett{k})$ has the following expression:
\beq
\widetilde{\Omega}(\vett{k})=\omega'k_z-\frac{\omega''}{2}k_z^2+\frac{\omega'}{2k}k_{\perp}^2.
\eeq
This can be simply checked by substituting Eq. \eqref{eq11} into Eq. \eqref{eq6} and solving for $\widetilde\Omega(\vett{k})$. Moreover, the above result allows us to introduce the co-moving coordinate $Z=z-\omega't$ and the new frequency $\Omega=-(\omega''k_z^2)/2+(\omega'k_{\perp}^2)/(2k_0)$ such that $k_zz-\widetilde\Omega(\vett{k})t=k_zZ-\Omega t$. If we now introduce the normalized transverse wave vector $\alpha$ and the X wave velocity $v$ \cite{comment1} as
\beq
\left\{\begin{array}{ll}
k_{\perp}=\alpha\sqrt{\frac{\omega''k}{\omega'}},\\
k_z=\alpha-\frac{v}{\omega''},
\end{array}\right.
\eeq
Eq.\eqref{eq11} can be rewritten in the following form:
\barr\label{eq12}
\fieldA{r}{t}&=&\sum_m\,d_m\,e^{im\theta}\,\int\,dv\,\int_0^{\infty}\,d\alpha\, X(\alpha,v)\nonumber\\
&\times&J_m\left(\sqrt{\frac{\omega''k}{\omega'}}\alpha R\right)e^{i\left(\alpha-\frac{v}{\omega''}\right)(Z-vt)}\nonumber\\
&\times&e^{-i\frac{v^2t}{2\omega''}},
\earr
where
\beq
X(\alpha,v)=\frac{k\alpha}{\omega'}\,S\left(\sqrt{\frac{\omega''k}{\omega'}}\alpha,\alpha-\frac{v}{\omega''}\right),
\eeq
is the so called X wave transform \cite{XwaveTransform}. A suitable choice for the spectrum $X(\alpha,v)$ is represented by the generalised X wave spectrum presented in Ref. \cite{orthX}, namely
\beq
X(\alpha,v)=\sum_{p=0}^{\infty}\tilde{C}_p(v)f_p(\alpha),
\eeq
where $\tilde{C}_p(v)$ are some arbitrarily expansion coefficients and 
\beq\label{Eq_fp}
f_p(\alpha)=\sqrt{\frac{k\Delta}{\pi^2\omega'(1+p)}}(\alpha)L_p^{(1)}(2\alpha\Delta)e^{-\alpha\Delta},
\eeq
being $L_p^{(1)}$ the generalised Laguerre functions of the first kind of index $p$ \cite{nist} and $\Delta$ is a normalisation length related to the spatial extension of the spectrum. Substituting this result into Eq. \eqref{eq12} and introducing the generalised OAM-carrying X wave of order $p$ and velocity $v$ as
\barr\label{eq13}
\fieldX{m}{p}{v}&=&\int_0^{\infty}\,d\alpha\,f_p(\alpha)\,J_m\left(\sqrt{\frac{\omega''k}{\omega'}}\alpha R\right)\nonumber\\
&\times&e^{i\left(\alpha-\frac{v}{\omega''}\right)\zeta}\,e^{im\theta},
\earr
where $R=\sqrt{x^2+y^2}$, $\zeta=z-vt$ is the usual coming coordinate associated to X waves \cite{localizedWaves} and $C_{m,p}(v)\equiv\,d_m\tilde{C}_p(v)$, we can rewrite Eq. \eqref{eq12} in the following, inspiring, form:
\beq\label{eq14}
\fieldA{r}{t}=\sum_{m,p}\,\int\,dv\,C_{m,p}(v)e^{-\frac{iv^2t}{2\omega''}}\fieldX{m}{p}{v}.
\eeq
This is the first result of our work. The propagation of a paraxial electromagnetic field in a dispersive medium can be described by expanding the field envelope as a superposition of generalised  X waves of order $p$ and velocity $v$, carrying $m$ units of OAM. To emphasise this, in the above equations we introduce the notation $\fieldX{m}{p}{v}$ to highlight the three degrees of freedom that later upon quantisation will be promoted to quantum nnumbers, namely the OAM $m$, the spectral order $p$ and the X wave velocity $v$. Note, moreover, that while $m$ and $p$ are discrete quantum numbers (i.e., the correspondent operators have discete spectra), the X wave velocity is a continuous quantum number. This result, moreover, is corroborated by the fact that  $\fieldX{m}{p}{v}$ are a complete and orthogonal set of functions and can be therefore used as a basis to represent any arbitrary field envelope $\fieldA{r}{t}$. A proof of the orthogonally and completeness of generalised OAM-carrying X waves is given in Appendix A.
\subsection{Total Energy of the Field}
In preparation for the quantisation of the field $\fieldA{r}{t}$, it is instructive to calculate the total energy carried by the field $\fieldE{r}{t}$ or, equivalently, the Hamiltonian function corresponding to $\fieldE{r}{t}$. As already pointed out in Eq. \eqref{eq2}, the total energy $\mathcal{E}$ carried by the field is simply obtained by integrating the quantity $|\fieldA{r}{t}|^2$ over the whole space. If we use the expression for $\fieldA{r}{t}$ given by Eq. \eqref{eq14} and substitute it into Eq. \eqref{eq3}, it is not difficult to show that $\mathcal{E}$ has the following well known expression \cite{jackson}:
\beq\label{eqEnergy}
\mathcal{E}=\int\,d^3r\,|\fieldA{r}{t}|^2=\sum_{p,m}\,\int\,dv\,|\,C_{m,p}(v)\,|^2.
\eeq
We can rewrite the above expression in a more inspiring form by noticing that we can define the following time-dependent expansion coefficients from Eq. \eqref{eq14}, namely
\beq
\mathcal{C}_{m,p}(v,t)=C_{m,p}(v)e^{-i\frac{v^2}{2\omega''}t},
\eeq
and observe that they are solution to the harmonic oscillator differential equation, i.e., 
\beq
\frac{d^2\mathcal{C}_{m,p}(v,t)}{dt^2}=-\omega_{m,p}(v)^2\mathcal{C}_{m,p}(v,t),
\eeq
with
\beq\label{defOmega}
\omega_{m,p}(v)=\frac{v^2}{2\omega''}.
\eeq
With this result at hand, we can rewrite Eq. \eqref{eq14} as
\beq\label{eq15}
\fieldA{r}{t}=\sum_{p,m}\,\int\,dv\,\mathcal{C}_{m,p}(v,t)\,\fieldX{m}{p}{v}.
\eeq
If we then notice that $|C_{m,p}(v)|^2=|\mathcal{C}_{m,p}(v,t)|^2$, we can also rewrite the total energy of the field as
\beq\label{eq16}
\mathcal{E}=\sum_{p,m}\,\int\,dv\,|\,\mathcal{C}_{m,p}(v,t)\,|^2.
\eeq
Equations \eqref{eq15} and \eqref{eq16} have a very simple interpretation: the field envelope $\fieldA{r}{t}$ (and, consequently, the electric field itself) can be viewed as a collection (integral sum) of harmonic oscillators, each of them with complex amplitude $C_{m,p}(v)$ and associated to a travelling invariant wave with a $v$-dependent resonant frequency $\omega_{m,p}(v)$, corresponding to the kinetic energy of free particles.

The result above is similar to the case of a field expanded onto the normal modes of an optical cavity, with the different that in this case the modes are continuously distributed and rigidly moving, instead of being standing waves as in the traditional case. This allows us to adapt the ordinary quantisation techniques, as the one described for example in Ref. \cite{loudon}, to the case of X waves in dispersive media.
\section{Quantisation of X waves}
To quantise the field given by Eq. \eqref{eq15}, we employ the standard technique of expressing the total energy $\mathcal{E}$ as a collection of harmonic oscillators and then associate creation and annihilation operators to the field itself \cite{loudon}. As discussed in the previous section, according to Eq. \eqref{eq16} the total energy of the field is already written as a (continuous) collection of harmonic oscillators, and therefore the quantisation is immediate. To make this more explicit, however, we introduce the two real quantities $Q_{m,p}(v)$ and $P_{m,p}(v)$ and write the complex amplitudes $C_{m,p}(v)$ as
\beq\label{quant0}
C_{m,p}(v)=\frac{1}{\sqrt{2}}\left[\omega_{m,p}(v)Q_{m,p}(v)+iP_{m,p}(v)\right],
\eeq
such that
\beq
|C_{m,p}(v)|^2=\frac{1}{2}\left[P^2_{m,p}(v)+\omega_{m,p}^2(v)Q^2_{m,p}(v)\right].
\eeq
It is therefore not difficult to interpret $Q_{m,p}(v)$ and $P_{m,p}(v)$ as the position and momentum of the field. We can then promote these quantities to operators and introduce the creation and annihilation operators of the field in the traditional way as follows:
\bseq\label{quant1}
\begin{align}
\hat{Q}_{m,p}(v)&=\sqrt{\frac{\hbar}{2\omega_{m,p}(v)}}\left[\creatX{m}{p}{v}+\annX{m}{p}{v}\right],\\
\hat{P}_{m,p}(v)&=i\sqrt{\frac{\hbar\omega_{m,p}(v)}{2}}\left[\creatX{m}{p}{v}-\annX{m}{p}{v}\right],
\end{align}
\eseq
with the usual bosonic commutation relations \cite{note24}
\bseq
\begin{align}
\left[\annX{m}{p}{v},\creatX{n}{q}{u}\right]&=\delta_{m,n}\delta_{p,q}\delta(u-v),\\
\left[\annX{m}{p}{v},\annX{n}{q}{u}\right]&=0=\left[\creatX{m}{p}{v},\creatX{n}{q}{u}\right].
\end{align}
\eseq
Substituting Eqs. \eqref{quant0} and \eqref{quant1} into Eq. \eqref{eq15} brings to the following expression for the quantised field
\barr\label{fieldOp}
\annA{r}{t}&=&\sum_{m,p}\,\int\,dv\,e^{-\frac{i}{\hbar}\left(\frac{Mv^2}{2}\right)t}\sqrt{\hbar\omega_{m,p}(v)}\nonumber\\
&\times&\psi_{m,p}^{(v)}(\vett{r},\zeta)\annX{m}{p}{v}+\text{h.c.},
\earr
where $\text{h.c.}$ stands for hermitian conjugate. The expression for the Hamilton operator can be instead found by substituting Eqs. \eqref{quant0} and \eqref{quant1} into Eq. \eqref{eq16}, thus obtaining
\beq\label{hamQ}
\hat{H}=\sum_{m,p}\,\int\,dv\,\hbar\omega_{m,p}(v)\left\{\creatX{m}{p}{v}\annX{m}{p}{v}+\frac{1}{2}\right\}.
\eeq
Moreover, if we introduce the ``mass of the X wave" as $M\equiv\hbar/\omega''$, the above equation can be rewritten as follows:
\beq\label{ham2}
\hat{H}=\sum_{m,p}\,\int\,dv\,\frac{Mv^2}{2}\left\{\creatX{m}{p}{v}\annX{m}{p}{v}+\frac{1}{2}\right\}.
\eeq
Equations \eqref{fieldOp}-\eqref{ham2} are the second main result of our work and represent quantisation of generalised OAM-carrying X waves. In particular, Eq. \eqref{ham2} is the analogue of a quantised Hamiltonian of a 1D gas of weakly interacting bosons, with mass $M$ and velocity $v$ \cite{1Dgas}. This analogy is quite important, as it allows us to treat quantum X waves, which are intrinsically 3+1-dimensional fields, as one dimensional particle-like objects parametrised only by their own velocities $v$. This feature is similar to the particle-like nature of quantum solitons \cite{solitons} and it is ultimately due to the nondiffracting nature of X waves.
\subsection{Particle States}
We can now look for the eigenstates of the Hamilton operator, represented by either Eq. \eqref{hamQ} or \eqref{ham2}. In particular, in analogy to the traditional case \cite{loudon}, we can introduce the $N$- particle states $\ket{m,p,v,N}$ as states containing $N$ field excitation with velocity $v$ in the travelling mode (i.e., the X wave) $\psi_{m,p}^{(v)}(\vett{r},t)$ with energy $Mv^2/2$. These states are mutually orthogonal
\beq\label{FockNorm}
\braket{n,q,u,N}{m,p,v,N}=\delta_{m,n}\delta_{p,q}\delta(u-v),
\eeq
complete and they can be obtained from the vacuum state $\ket{0}$ by successive applications of the creation operator, i.e.,
\beq
\ket{m,p,v,N}=\frac{1}{\sqrt{N!}}\left[\creatX{m}{p}{v}\right]^N\ket{0}.
\eeq
Moreover, it is not difficult to prove that the usual results concerning Fock states \cite{loudon} are also valid in this case, namely the expectation value of the electric field operator is zero on the particle eigenstates, i.e., 
\beq
\expectation{m,p,v,N}{\annA{r}{t}}{m,p,v,N}=0.
\eeq
However, the Fock states $\ket{m,p, v,N}$ are not normalisable, since their representation in configuration state $\braket{\vett{r},t}{m,p,v,N}=\psi_{m,p}^{(v)}(\vett{r},t)$ carries infinite energy. In fact, if we compute the expectation value of the Hamilton operator over the Fock states, we get the following result \cite{comment2}
\beq\label{eq36}
\expectation{m,p,v,N}{\hat{H}}{m,p,v,N}=\frac{NMv^2}{2}\delta(v-v).
\eeq
The Dirac delta appearing above, comes from the normalisation condition \eqref{FockNorm}. This is not surprising, since also in the classical case, X waves are non-normalisable solutions of Maxwell's equations (exactly as plane waves), due to the fact that they carry infinite energy. 

To solve this issue, there are different approaches possible. First, one could introduce a finite quantisation volume and carry out the field quantisation is such finite volume \cite{loudon}. In doing this, however, one should be careful to introduce a finite cavity with the desired symmetry, to use X waves as eigenstates for the field expansion. Another possibility would be to redefine the normalisation condition given by Eq. \eqref{FockNorm} by imposing by hand that it only makes sense when calculated for X states with the same velocity, namely
\beq
\braket{n,q,v,N}{m,p,v,N}\equiv\delta_{m,n}\delta_{p,q}.
\eeq
In this case, however, all the observable quantities will be finite if defined per unit volume, while the integrated quantities will be infinite, as in the standard case for X waves \cite{localizedWaves}. In particular, the Hamiltonian density in this case would be simply given by
\beq
\hat{\mathcal{H}}=\hat{A}^{\dagger}(\vett{r},t)\hat{A}(\vett{r},t),
\eeq
and its expectation value over Fock states $\langle\hat{\mathcal{H}}\rangle\equiv\expectation{m,p,v,N}{\hat{\mathcal{H}}}{m,p,v,N}$ will be
\beq
\langle\hat{\mathcal{H}}\rangle=N\hbar\omega_{m,p}(v)|\psi_{m,p}^{(v)}(\vett{r},\zeta)|^2,
\eeq
which is finite. However,
\beq
\int\,d^3r\,\langle\hat{\mathcal{H}}\rangle=N\hbar\omega_{m,p}(v)\int\,d^3r\,|\psi_{m,p}^{(v)}(\vett{r},\zeta)|^2=\infty,
\eeq
since X waves carry infinite energy. Moreover, in analogy with what is done for monochromatic paraxial beams \cite{haus}, all the other observables can be normalised using the energy density, thus obtaining finite, well defined quantities per unit volume.

Another possible way to overcome the problem in Eq. \eqref{eq36} is to consider X waves with finite energy (such as Bessel-X pulses, for example \cite{localizedWaves}). In this case, we can define a new set of Fock states (to differentiate them from the set of Fock states corresponding to infinite-energy X waves) $\ket{X_{m,p};n}$, which are normalisable, since 
\beq
\int\,d^3r\, |X_{m,p}(\vett{r},t)|^2<\infty,
\eeq
where $X_{m,p}(\vett{r},t)=\braket{\vett{r},t}{X_{m,p};n}$. We therefore have
\beq
\braket{X_{n,q};k}{X_{m,p};r}=\delta_{n,m}\delta{p,q}\delta_{k,r},
\eeq
and the expectation value of $\hat{H}$ over such states gives now a finite value.

Having clarified this point, in the remaining of the manuscript, we will actually employ (unless otherwise explicitly specified) Eq. \eqref{FockNorm} as ``normalisation condition" for the X waves particle states, implicitly remembering that they are associated with waves carrying infinite energy.	
\subsection{Coherent States}
In analogy with the case of single mode quantum optics \cite{loudon}, we can also define coherent X wave states as follows:
\beq\label{eqCoherent}
\ket{\alpha_{m,p}(v)}=e^{-\frac{|\alpha_{m,p}(v)|^2}{2}}\sum_{n=0}^{\infty}\frac{\alpha_{m,p}(v)^n}{\sqrt{n!}}\ket{m,p,v,n},
\eeq
being $\alpha_{m,p}(v)\in\mathbb{C}$. Then, the expectation value of the field operator $\annA{r}{t}$ gives the classical X wave field, i.e., 
\beq
\expectation{\alpha_{m,p}(v)}{\annA{r}{t}}{\alpha_{m,p}(v)}=\mathcal{N}\psi_{m,p}^{(v)}(\vett{r},\zeta),
\eeq
which, apart from the normalisation factor
\beq
\mathcal{N}=\alpha_{m,p}(v)\sqrt{\hbar\omega_{m,p}(v)}e^{-\frac{i}{\hbar}\left(\frac{Mv^2}{2}\right)t},
\eeq
corresponds to the classical generalised OAM-carrying X-wave field introduced in Eq. \eqref{eq13}. Like their classical counterparts, coherent X wave states carry infinite energy, since
\beq\label{cohHX}
\expectation{\alpha_{m,p}(v)}{\hat{H}}{\alpha_{m,p}(v)}=\frac{Mv^2}{2}|\alpha_{m,p}(v)|^2\delta(v-v).
\eeq

%
%
\section{Second Order Nonlinearity}
We now turn our attention to quantum nonlinear processes involving X waves, and in particular to $\chi^{(2)}$-processes such as optical parametric amplification and Kerr nonlinearity. To start with, let us then consider three fields, namely a pump field, which, for the sake of simplicity, will be treated as a bright coherent state, characterised by the frequency $\omega_p$, and a signal and idler fields characterised by the frequencies $\omega_1$ and $\omega_2$, respectively. Moreover, we assume that signal and idler have different group velocities, i.e., $\omega_1'\neq\omega_2'$, but the group velocity dispersion of both fields is the same, namely $\omega_1''=\omega_2''\equiv\omega''$. The field operators for the signal and idler fields are obtained from Eq. \eqref{fieldOp} with the substitutions $\omega\rightarrow\omega_{1,2}$ and $\zeta\rightarrow\zeta_{1,2}$, respectively. The total (time dependent)Hamiltonian of the system in presence of second order nonlinearity is then given as follows:
\beq\label{nonlinearH}
\hat{H}(t)=\hat{H}_0(t)+\lambda\hat{H}_I(t),
\eeq
where $\hat{H}_0$ is the free field Hamiltonian given by Eq. \eqref{hamQ}, $\lambda\ll 1$ is the interaction parameter (assumed very small) and $\hat{H}_I$ is the quantised version of the classical $\chi^{(2)}$-interaction Hamiltonian \cite{boyd}, namely
\beq\label{eq48}
\mathcal{H}_I=\chi\braket{A_1}{A_2^*}+\chi^*\braket{A_2}{A_1^*}.
\eeq
Before considering on the nonlinear quantum dynamics of the system described by Eq. \eqref{nonlinearH}, we must find a quantised expression for the classical interaction Hamiltonian. To do so, let us consider the first term in the above equation (the second is then obtained by simply taking the Hermitian conjugate) and uses Eq. \eqref{fieldOp} to obtain
\barr\label{eq49}
&&\chi\braket{A_1}{A_2^*}=\chi\int\,d^3r\,\hat{A}^{\dagger}_1(\vett{r},t)\hat{A}^{\dagger}_2(\vett{r},t)\nonumber\\
&=&\chi\sum_{m,p,n,q}\int\,du\,dv\,e^{\frac{i(u^2+v^2)t}{2\omega''}}\sqrt{\hbar^2\omega_{m,p}(u)\omega_{n,q}(v)}\nonumber\\
&\times&\int\,d^3r\,\psi_{m,p,1}^{(u)*}(\vett{r},\zeta_1)\psi_{n,q,2}^{(v))}(\vett{r},\zeta_2)\,\creatX{m}{p}{u}\creatXb{n}{q}{v},
\earr
where the subscript $1$ refers to the signal field and $2$ to the idler field, respectively. We can now use the above result to write the quantised form of the interaction Hamiltonian for second order nonlinearity. To do this, however, it is useful to define the quantity $\rho=\sqrt{k_1\omega_2'/(k_2\omega_1')}$ and introduce the interaction function
\barr
\chi_{mpq}(x)&=&\frac{(-1)^m4\pi^2\rho\omega_1'}{k_1x\sqrt{\omega_2'}}\,f_{p,1}\left(\frac{x}{\omega"(1+\rho)}\right)\nonumber\\
&\times&f_{q,2}\left(\frac{\rho x}{\omega"(1+\rho)}\right)\Theta(x),
\earr
where $\Theta(x)$ is the Heaviside step function \cite{nist} and $f_{p,1}$, $f_{q,2}$ are defined according to Eq. \eqref{Eq_fp}. Moreover, we also define the function
\beq
F(u,v)=\frac{u^2+v^2}{2\omega"}+\frac{(v-\rho u)(u-v+\omega_1'-\omega_2')}{\omega"(1+\rho)}.
\eeq
If we now subsitute Eq. \eqref{eq49} (and its complex conjugate) into Eq.\eqref{eq48} and perform the spatial integration using the orthogonality relation between X waves modes given by Eq. \eqref{EqOrtho}, we can write the interaction Hamiltonian in the following form:
\barr\label{intHam}
\hat{H}_I&=&\hbar\sum_{m,p, q}\,\int\,du\,dv\,\chi_{m,p, q}(u+v)\nonumber\\
&\times&\sqrt{\omega_{m,p}(u)\omega_{-m,q}(v)}e^{i F(u,v) t}\creatX{m}{p}{u}\creatXb{-m}{q}{v}\nonumber\\
&+&\text{h.c.}
\earr 

We are now in the position to calculate the state of the electromagnetic field after the nonlinear interaction with the medium. To do so, since $\lambda\ll1$, we can treat the nonlinearity as perturbation and use the Schwinger-Dyson expansion of the propagator $\exp{[-i\hat{H}(t)/\hbar]}$ \cite{messiah} truncated at the first order to obtain the following expression for the state of the field after the nonlinear interaction:
\beq
\ket{\psi^{(1)}(t)}=-\frac{i}{\hbar}\int_0^t\,d\tau\,\hat{H}_I(\tau)\ket{0},
\eeq
where $\ket{\psi(0)}=\ket{0}$ has been implicitly assumed. In our case, using the expression of the interaction Hamiltonian given by Eq. \eqref{intHam} we have
\barr
&-&\frac{i}{\hbar}\int_0^t\,d\tau\,\hat{H}_I(\tau)=-i\sum_{m,p,q}\,\int\,du\,dv\chi_{m,p,q}(u+v)\nonumber\\
&\times&\sqrt{\omega_{m,p}(u)\omega_{-m,q}(v)}\int_0^t\,d\tau\,e^{iF(u,v)\tau}\,\creatX{m}{p}{u}\creatXb{-m}{q}{v}.\nonumber
\earr
If we now introduce the quantities
\barr
K(u,v)&=&\frac{(v-\rho u)(u-v+\omega_1'-\omega_2')}{2\omega''(1+\rho)},\\
G(u,v,t)&=&-\frac{2i}{F(u,v)}\sin\left(\frac{F(u,v)t}{2}\right),
\earr
and introduce, for convenience of notation, the quantity
\barr
\mathcal{G}_{m,p,q}(u,v,t)&=&\sqrt{\omega_{m,p}(u)\omega_{-m,q}(v)}G(u,v,t)\nonumber\\
&\times&e^{iK(u,v)t}\chi_{m,p,q}(u+v),
\earr
and the two particle state
\beq
|m,p,u;-m,q,v\rangle\equiv\creatX{m}{p}{u}\creatXb{-m}{q}{v}\ket{0},
\eeq
we can write the perturbed state $\ket{\psi^{(1)}(t)}$ as follows:
\beq\label{CVent}
\ket{\psi^{(1)}(t)}=\sum_{m,p,q}\,\int\,du\,dv\,\mathcal{G}_{m,p,q}(u,v,t)|m,p,u;-m,q,v\rangle.
\eeq
The above state represents the superposition of two particles, corresponding to the two modes $\omega_1$ and $\omega_2$ (generated by the nonlinear process) travelling with velocities $u$ and $v$, respectively. Moreover, since the function $\mathcal{G}_{m,p, q}(u,v,t)$ is in general non-separable in the variables $u$ and $v$, the above state represents a continuous variable entangled state in the (continuous valued) velocities $u$ and $v$ of the two particles.
\subsection{Transition Probability}
We can now use the explicit, analytic expression for the state after the nonlinear interaction to calculate the probability for the field to be in such a state after the interaction with the $\chi^{(2)}$-nonlinearity of the medium. We then get
\begin{widetext}
\barr\label{transProb}
P(t)&=&\sum_{m,p,q}\sum_{n,r,s}\int\,\,dudvd\tilde{u}d\tilde{v}\,\mathcal{G}^*_{n,r,s}(\tilde{u},\tilde{v},t)\mathcal{G}_{m,p,q}(u,v,t)\braket{n,r,\tilde{u};-n,s,\tilde{v}}{m,p,u;-m,q,v}\nonumber\\
&=&\sum_{mp,pq}\,\int\,du\,dv\,\mathcal{P}_{m,p,q}(u,v,t),
\earr
\end{widetext}
where
\barr
\mathcal{P}_{m,p,q}(u,v,t)&=&\omega_{m,p}(u)\omega_{-m,q}(v)|\chi_{m,p,q}(u+v)|^2\nonumber\\
&\times&\text{sinc}^2\left(\frac{F(u,v)t}{2}\right)t^2.
\earr
As it can be seen, the transition probability is ,in general, non-separable in $\{u,v\}$ due to the inseparability of 
$\chi_{m,p,q}(u+v)$ with respect to $u$ and $v$. Although the above form represents an exact (within the perturbative limit) expression for the transition probability $P(t)$, its form is quite difficult to handle, and further approximations are needed to fully understand its properties. In particular, here we employ two different approximations. First, we consider the state of the system for $t\rightarrow\infty$, meaning that we look at the system far away (in time) from the moment of interaction. If we do that, $\mathcal{P}_{m,p,q}(u,v,t)$ tends to a Dirac delta function peaked at $F(u,v)=0$, since
\beq
\lim_{t\rightarrow\infty}\text{sinc}^2\left[\frac{F(u,v)t}{2}\right]=\delta(F(u,v)).
\eeq
We then have
\barr\label{ProbDens}
\mathcal{P}_{m,p,q}(u,v; t\rightarrow\infty)&=&\omega_{m,p}(u)\omega_{-m,q}(v)|\chi_{m,p,q}(u+v)|^2\nonumber\\
&\times&\delta(F(u,v))t^2.
\earr
For large times, therefore, the entangled particles with velocities $u$ and $v$ are associated to points in the $(u,v)-$plane which are constrained on the parabolic surface $F(u,v)=0$.

To further simplify this result, we can look at the so-called low velocity limit, which corresponds, in the theory of interacting quantum gases \cite{quantumGas}, to the small momentum approximation. In this limit, we neglect the quadratic contributions in $F(u,v)$. This allows us to rewrite the function $F(u,v)$ as follows
\beq
F(u,v)\simeq\,\frac{\omega_1'-\omega_2'}{2\omega''(1+\rho)}(v-\rho u),
\eeq
and therefore the Dirac delta above can be rewritten as
\beq
\delta(F(u,v))\simeq\,\frac{2\omega''(1+\rho)}{|\omega_1'-\omega_2'|}\delta(v-\rho u),
\eeq
Inserting this approximation in Eq. \eqref{ProbDens} brings to the following result
\barr
\mathcal{P}_{m,p,q}(u,v;\infty)&\simeq&\,\omega_{m,p}(u)\omega_{-m,q}(v)|\chi_{m,p,q}(u+v)|^2\nonumber\\
&\times&\frac{2\omega''(1+\rho)}{|\omega_1'-\omega_2'|}\delta(v-\rho u)t^2.
\earr
For practical cases, $\rho\simeq 1$ and therefore the above expression constrain the two particles to travel at the same speed $u=v$. In this case, the transition probability \eqref{transProb} assumes the following explicit form
\barr
P(t)&=&\frac{4\omega''t^2}{|\omega_1'-\omega_2'|}\sum_{mp,pq}\,\int\,du\,\omega_{m,p}(u)\omega_{-m,q}(u)\nonumber\\
&\times&\left|\chi_{m,p, q}(2u)\right|^2.
\earr
The above integral over $u$ can be performed analytically if we use the definition of generalised Laguerre polynomials \cite{ref27pre} and the integral formula \cite{gradstein}
\beq
\int_0^{\infty}\,dx\,x^k\,e^{-\beta x}=\frac{k!}{\beta^{k+1}},
\eeq
thus obtaining
\barr
\mathcal{C}_{p,q}&=&\int_0^{\infty}\,du\,\omega_{m,p}(u)\omega_{-m,q}(u)\,\left|\chi_{m,p, q}(2u)\right|^2\nonumber\\
&=&\left[\frac{\pi^2}{\omega_2'(1+p)(1+q)}\right]\sum_{s=0}^p\sum_{t=0}^q\left[\binom{p+1}{p-s}\binom{q+1}{q-t}\right]^2\nonumber\\
&\times&\frac{(2s+2t+1)!}{4^{s+t+2}(s!t!)^2}.
\earr
The transition probability in the large times, low velocity limit thus reads (with $C\equiv\sum_{p,q}\mathcal{C}_{p,q}$)
\beq
P(t)=\left(\frac{4C\omega''}{|\omega_1'-\omega_2'|}\right)t^2.
\eeq
Therefore, in the large times and low velocity limit, the transition probability scales with the square of the interaction time. This result, anyway, is not unexpected, as in our case the large times limit corresponds to the infinite thickness, perfect phase-matching limit for standard quantum optics in bulk crystals \cite{boyd}.
\section{Kerr Effect}
We now turn our attention to the case of Kerr nonlinearity and study the dynamics of OAM-carrying X waves in presence of such nonlinearity. To do that, we first derive, in the low velocity limit, an expression for the interaction Hamiltonian and then calculate the state of the system after the interaction, assuming that the initial state of the field is the vacuum state. Afterwards, we show that the Kerr dynamics can be split into two parts: a classical evolution of the envelope function, which obeys a wave equation, and a quantum evolution for the field operators, which is described by a nonlinear Schr\"odinger equation, whose potential depends on the OAM content of the system.
\subsection{Interaction Hamiltonian}
The quantised form of the interaction Hamiltonian in the case of Kerr nonlinearity can be derived from its classical counterpart \cite{boyd} and reads as follows:
\beq
\hat{H}_I=\frac{\chi}{2}\int\,d^3r\creatA{r}{t}\creatA{r}{t}\annA{r}{t}\annA{r}{t}.
\eeq
If we substitute the expressions for the field operators as given by Eq. \eqref{fieldOp} and its Hermitian conjugate,  the above equation can be written as follows:
\barr\label{intKerr1}
\hat{H}_I&=&\frac{\hbar^2\chi}{2}\sum_{\{\vett{m}\}}\int d\vett{u}\, d^3r\,e^{\frac{im}{\hbar^2}(u^2+v^2-w^2-y^2)t}\nonumber\\
&\times&\mathcal{S}_{\vett{m}}(\vett{u},\vett{R},\zeta)\creatX{m}{p}{u}\creatX{n}{q}{v}\annX{l}{r}{w}\annX{s}{t}{y},
\earr
where we have introduced the shorthand notation $\{\vett{m}\}=\{m,n,l,s,p,q,r,t\}$ (with $\{m,n,l,s\}\in (-\infty, \infty)$ and $\{p,q,r,t\}\in[0,\infty)$) and $\vett{u}=\{u,v,\tilde{u},\tilde{v}\}$. The quantity $\mathcal{S}_{\vett{m}}(\vett{u},\vett{R},\zeta)$ is defined as 
$\mathcal{S}_{\vett{m}}(\vett{u},\vett{R},\zeta)=\mathcal{S}_0\,\fieldXc{m}{p}{u}\fieldXc{n}{q}{v}\fieldX{l}{r}{\tilde{u}}\fieldX{s}{t}{\tilde{v}}$, where $\mathcal{S}_0=\sqrt{\omega_{mp}(u)\omega_{nq}(v)\omega_{lr}(\tilde{u})\omega_{st}(\tilde{v})}$. If we now assume $t=0$, perform the spatial integration and introduce the vertex function $\chi_{\multi}(\eta)$  (see Appendix B for the details), after some manipulation we can rewrite the above interaction Hamiltonian in the following compact form:
\barr\label{Hexact}
\hat{H}_I&=&\frac{1}{2}\sum_{\{\vett{m}\}}\int d^4v\,\chi_{\{\vett{m}\}}(v_3-v_4+v_1+v_2)\nonumber\\
&\times&\sqrt{\omega_{mp}(v_1)\omega_{nq}(v_2)\omega_{lr}(v_3)\omega_{st}(v_4)}\nonumber\\
&\times&\creatX{m}{p}{v_1}\creatX{n}{q}{v_2}\annX{l}{r}{v_3}\annX{s}{t}{v_4},
\earr
where $\chi_{\{\vett{m}\}}(X)$ is the vertex function, as defined in appendix B.

Although the above expression for the Kerr Hamiltonian is exact, it cannot be treated analytically any further, thus limiting the amount of insight one can get about the effect of Kerr nonlinearity on OAM-carrying X waves. To this aim, we now introduce the so-called low velocity approximation, which corresponds to neglect quadratic terms in the expression of the eigenfrequencies $\omega_{ik}(v)$\cite{nota2}. Moreover, we introduce the Fourier transform of the vertex function (see Appendix B)
\beq
\chi_{\multi}(P)=\int d\eta\sigma_{\multi}(\eta)e^{-i\frac{P\eta}{\omega''}},
\eeq
and the Fourier representation for the creation and annihilation operators
\bseq\label{opFourier}
\begin{align}
\annX{m}{p}{v}&=\int \,d\eta\,\annF{m}{p}{\eta}e^{iv\eta/\omega''},\\
\annF{m}{p}{\eta}&=\frac{1}{2\pi\omega'}\int \,dv\,\annX{m}{p}{v}e^{-iv\eta/\omega''}.
\end{align}
\eseq
Using the quantities above, the Kerr Hamiltonian in the low velocity limit (and for $t=0$) assumes the following form:
\begin{widetext}
\beq\label{eq70}
\hat{H}_I=\sum_{\multi}\delta_{m+n,l+s}\int \,d\eta\,\Sigma_{\multi}(\eta)\creatF{m}{p}{\eta}\creatF{n}{q}{\eta}\annF{l}{r}{\eta}\annF{s}{t}{\eta}.
\eeq
\end{widetext}
The explicit expression of $\Sigma_{\multi}(\eta)$ (which essentially comprises $\sigma_{\multi}(\eta)$ and the result of the integration over $d^4v$) is given in Appendix C. A closer inspection to the above form of the Kerr Hamiltonian reveals that it has the same form as the standard quantum Kerr Hamiltonian used, for example, to derive quantum solitons in optical fibers \cite{solitons}.
\subsection{Time Evolution}
The evolution of a system under the action of the interaction Hamiltonian described by Eq. \eqref{eq70} can be easily calculated in the Heisenberg representation, where the time evolution is applied to the field operators, while the state does not evolve in time \cite{messiah}. First, we write the field operators \eqref{opFourier}  as
\bseq
\begin{align}
\annX{m}{p}{v}&\rightarrow\annXt{m}{p}{v}=\annX{m}{p}{v}e^{-i\omega_{mp}(v)t},\\
\annF{m}{p}{\eta}&\rightarrow\annFt{m}{p}{\eta}=\annF{m}{p}{\eta}e^{i\omega_{mp}(v)t}.
\end{align}
\eseq
Then, we use the above relations to rewrite the field operators $\creatA{r}{t}$ and $\annA{r}{t}$ as a function of $\creatFt{m}{p}{\eta}$ and $\annFt{m}{p}{\eta}$, thus obtaining
\beq\label{eq72}
\annA{r}{t}=\sum_{m,p}\int\,d\eta\,\xi_{mp}(\eta,\vett{r},t)\annF{m}{p}{\eta},
\eeq
where
\beq
\xi_{mp}(\eta,\vett{r},t)=\int\,dv\sqrt{\hbar\omega_{mp}(v)}e^{i\frac{2\eta v-v^2}{2\omega''}}\fieldX{m}{p}{v}.
\eeq
Notice that the form of the field operator given by Eq. \eqref{eq72} is very similar to the one of a field operator of an optical beam, where the function $\xi_{mp}(\eta,\vett{r},t)$ plays the role of the mode function \cite{qoptics}. In particular, it is not difficult to see that the mode function $\xi_{mp}(\eta,\vett{r},t)$ is a solution of the initial wave equation Eq. \eqref{eq6}. This result is quite important, as it states that the mode function $\xi_{mp}(\eta,\vett{r},t)$ contains information only on the classical (i.e., deterministic)evolution of the system under the action of the Kerr nonlinearity and it is somehow decoupled from the quantum evolution, which only affects $\annF{m}{p}{\eta}$. 

To study the quantum evolution of the system, therefore, we impose that the operator $\annFt{m}{p}{\eta}$ obeys the Heisenberg equation of motion
\beq
i\frac{d\annFt{m}{p}{\eta}}{dt}=\left[\hat{H},\annFt{m}{p}{\eta}\right],
\eeq
where $\hat{H}=\hat{H}_0+\hat{H}_I$. Using a bit of algebra, it is not difficult to show that
\beq
\left[\hat{H}_0,\annF{m}{p}{\eta}\right]=-\pi\omega''^2\frac{\partial^2\annF{m}{p}{\eta}}{\partial\eta^2},
\eeq
and
\barr
\left[\hat{H}_i,\annF{m}{p}{\eta}\right]&=&-\sum_{\multid}\delta_{a+m,e+g}\Sigma_{\multit}(\eta)\nonumber\\
&\times&\creatF{a}{b}{\eta}\annF{e}{f}{\eta}\annF{g}{h}{\eta},
\earr
where $\multid=\{a,b,e,f,g,h\}$ and $\multit=\{a,m,e,g,b,p,f,h\}$, with $\{a,m,e,g\}\in]-\infty,\infty[$ and $\{b,p,f,h\}\in[0,\infty[$. The equation of motion for the field operator $\annF{m}{p}{\eta}$ can be then written in the following form:
\barr\label{systemSE}
i\hbar\frac{\partial\annFd{m}{p}}{\partial t}&=&-\frac{\hbar^2}{2M}\frac{\partial^2\annFd{m}{p}}{\partial\eta^2}\nonumber\\
&+&\sum_{\multid}V_{\multit}(\eta)\creatFd{a}{b}\annFd{e}{f}\annFd{g}{h},
\earr

where
\beq
V_{\multit}(\eta)=-\hbar\delta_{a+m,e+g}\Sigma_{\multit}(\eta).
\eeq
The field operator $\annFt{m}{p}{\eta}$ is then a solution of a set of coupled nonlinear Schr\"odinger equation. As can be seen, the coupling is essentially given by the coupling of the different OAM modes that define the nonlinear potential $V_{\multit}(\eta)$. This is the main effect of the Kerr nonlinearity, namely to introduce a coupling between different OAM values of different X wave states. This coupling, moreover, depends essentially from $\Sigma_{\multit}(\eta)$, i.e., from the overlap integral between the four modes involved in the nonlinear process.  The Kronecker delta in the definition of the nonlinear potential $V_{\multit}(\eta)$, moreover, does not fix univoquely a relation between the four OAM states involved in the process, but it imposes only the conservation of angular momentum between the states involved in the interaction by defining a family of possible sets of OAM states that can be generated via Kerr effect. This, in principle, could be used to generate single photon X wave states with high OAM content.
\section{Summary and Conclusions}
In this work, we have presented a rigorous theory of quantum X waves with orbital angular momentum in dispersive media. In particular, we have shown that quantised OAM-carrying X waves are formally analogue to a 1D gas of interacting bosons, characterised by a mass $M$ and a velocity $v$ [see Eq. \eqref{ham2}], which are, respectively, proportional to the group velocity dispersion and the Bessel cone angle of the X wave. We have then used these results to investigate the dynamics of quantum X waves in media exhibiting $\chi^{(2)}-$, as well as $\chi^{(3)}-$nonlinearities. For the case of quadratic nonlinearities, we have shown that a continuous variable entanglement between the X wave velocities can be realised [see Eq. \eqref{CVent}].  For the case of Kerr nonlinearity, instead, we have shown that the dynamics of the X wave can be splitted into a classical (deterministic) and a quantum part. The classical mode function $\xi_{mp}(\eta,\vett{r},t)$ evolves accordingly to the wave equation in dispersive media [Eq. \eqref{eq3bis}], while the quantum part evolves according to a system of coupled nonlinear Schr\"odinger equations [Eq. \eqref{systemSE}], with a potential that depends on the coupling between the various OAM states involved in the interaction.

In conclusion, our work presents a complete theoretical toolkit for the handling of nondiffracting quantum states of light and we envisage that it could be useful for the realisation of a new generation of quantum communication and quantum information protocols based on nondiffracting optical pulses. The natural resilience of X waves to external perturbations, in fact, makes them the idea candidate for the realisation of free space quantum communication channels. The fact that they carry OAM, moreover, gives the possibility to increase the amount of information that can be transferred in a (virtually) undistorted way through the atmosphere.

\section*{Acknowledgements}
C.C. acknowledges support from the Templeton foundation, grant number 58277. The authors thank the Deutsche Forschungsgemeinschaft (grants BL 547/13-1, SZ 276/9-1 and SZ 276/12-1) for financial support.

\section*{Appendix A: Orthogonality of Generalised OAM-carrying X Waves}
The aim of this appendix is to show that generalised OAM-carrying X waves of order $p$ and velocity $v$ $\fieldX{m}{p}{v}$ as given by Eq. \eqref{eq13}, constitute an orthogonal  set of functions. 

Let us consider the following scalar product between two generalised OAM-carrying X waves, one of order $p$ and velocity $v$, namely $\fieldX{m}{p}{v}$, and the other one of order $q$ and velocity $u$, i.e., $\fieldX{l}{q}{u}$
\beq
s\equiv\int\,d^3r\,\fieldX{l}{q}{u}^*\fieldX{m}{p}{v}.
\eeq
This quantity can be written explicitly by using Eq. \eqref{eq13} to obtain
\barr
s&=&\frac{k/(\pi\omega')}{\sqrt{(1+p)(1+q)}}\int\,dz\,\int_0^{\infty}\,d\alpha\,(\alpha\Delta)L_p^{(1)}(2\alpha\Delta)e^{-\alpha\Delta}\nonumber\\
&\times&\int_0^{\infty}\,d\beta\,(\beta\Delta)L_q^{(1)}(2\beta\Delta)e^{-\beta\Delta}\,\int_0^{2\pi}\,d\theta\,e^{i(m-l)\theta}\nonumber\\
&\times&\int_0^{\infty}\,dR\,R\,J_l\left(\sqrt{\frac{\omega''k}{\omega'}}\beta\,R\right)\,J_m\left(\sqrt{\frac{\omega''k}{\omega'}}\alpha\,R\right)\nonumber\\
&\times&e^{i\left[\left(\alpha-\frac{v}{\omega''}\right)(Z-vt)-\left(\beta-\frac{u}{\omega''}\right)(Z-ut)\right]}.
\earr
First, we solve the azimuthal integral, which gives
\beq
\int_0^{2\pi}\,d\theta\,e^{i(m-l)\theta}=2\pi\delta_{m,l}.
\eeq
Then, by substituting this result in the expression above and introducing the scaled radial coordinate $\rho=R\sqrt{\omega''k/\omega'}$, we have
\barr\label{appA1}
s&=&\left(\frac{2}{\omega''}\right)\frac{\Delta^2\delta_{m,l}}{\sqrt{(1+p)(1+q)}}\int dz\int_0^{\infty} d\alpha\int_0^{\infty} d\beta \,\alpha\,\beta e^{-\Delta(\alpha+\beta)}\nonumber\\
&\times&L_p^{(1)}(2\alpha\Delta)L_q^{(1)}(2\beta\Delta)e^{i\left[\left(\alpha-\frac{v}{\omega''}\right)(Z-vt)-\left(\beta-\frac{u}{\omega''}\right)(Z-ut)\right]}\nonumber\\
&\times&\int_0^{\infty}\,d\rho\,\rho\,J_m(\alpha\rho)\,J_m(\beta\rho).
\earr
We now use the orthogonality of Bessel functions \cite{nist}
\beq
\int_0^{\infty}\,d\rho\,\rho\,J_m(\alpha\rho)\,J_m(\beta\rho)=\frac{1}{\beta}\delta(\alpha-\beta),
\eeq
and we rewrite (using also the above result) the exponential function in Eq. \eqref{appA1} that contains $Z$ and $t$ as follows:
\barr
e^{i\left[\left(\alpha-\frac{v}{\omega''}\right)(Z-vt)-\left(\beta-\frac{u}{\omega''}\right)(Z-ut)\right]}&=&e^{-\alpha(v-u)t}e^{\frac{v^2-u^2}{2\omega''}t}\nonumber\\
&\times&e^{-i\frac{v-u}{\omega''}Z}.
\earr
Now, since $dz=dZ$, we can perform the $z$-integration in Eq. \eqref{appA1} obtaining
\beq
\int\,dZ\,e^{-i\frac{v-u}{\omega''}Z}=2\pi|\omega''|\delta(u-v).
\eeq
Substituting this result into Eq. \eqref{appA1} and using moreover the orthogonality relation for generalised Daguerre polynomials \cite{nist}
\beq
\int_0^{\infty}\,d\alpha\,\Delta\alpha L_p^{(1)}(2\alpha\Delta)L_q^{(1)}(2\alpha\Delta)e^{-2\Delta\alpha}=(p+1)\delta_{p,q},
\eeq
we obtain the final result
\beq
s=4\pi\Delta\text{sign}(\omega'')\delta_{m,l}\delta_{p,q}\delta(u-v).
\eeq
Apart from a normalisation constant [which can be included in the definition of $\fieldX{m}{p}{v}$], we then have that

\beq\label{EqOrtho}
\int\,d^3r\,\fieldX{l}{q}{u}^*\fieldX{m}{p}{v}=\delta_{m,l}\delta_{p,q}\delta(u-v).
\eeq
Then,  generalised OAM-carrying X waves are an orthogonal set. Moreover, from the above expression it can also be noted that generalised OAM-carrying X waves carry (unsurprisingly) infinite energy, as their norm is infinite, like for plane waves.
\section*{Appendix B: Kerr Hamiltonian and Vertex Function}
In this appendix, we calculate explicitly the spatial integral appearing in Eq. \eqref{intKerr1} and define the correspondent vertex function. In particular, we will first rewrite the integral appearing in Eq. \eqref{intKerr1} in simpler form, then define the vertex function and calculate its Fourier transform, which will allow us to give a compact insightful form to the vertex function  $\chi_{\{\vett{m}\}}(\vett{x})$. First of all, let us write the spatial integral explicitly (remembering that for $t=0$, $\zeta=z$):
\begin{widetext}
\barr
I&\equiv&\int d^3r\fieldXcz{m}{p}{u}\fieldXcz{n}{q}{v}\fieldXz{l}{r}{\tilde{u}}\fieldXz{s}{t}{\tilde{v}}=\int d^3r\int_0^{\infty}d^4\alpha f_p(\alpha_1)f_q(\alpha_2)f_r(\alpha_3)f_t(\alpha_4)\nonumber\\
&\times&J_m\left(\xi\alpha_1r\right)J_n\left(\xi\alpha_2r\right)J_l\left(\xi\alpha_3r\right)J_s\left(\xi\alpha_4r\right)e^{i(l+s-m-n)\theta}e^{i\left(-\alpha_1-\alpha_2+\alpha_3+\alpha_4+\frac{u+v-\tilde{u}-\tilde{v}}{\omega''}\right)z}\nonumber\\
&=&I(u+v-w-y)I_zI_{\theta},
\earr
\end{widetext}
where  $d\alpha^4=d\alpha_1d\alpha_2d\alpha_3d\alpha_4$ and $\xi=\sqrt{\omega''k/\omega'}$. Moreover, in the last line we used the notation $I_z$ and $I_{\theta}$ to indicate the integrals with respect to $z$ and $\theta$, which can be solved immediately, thus leading to

\beq
I_z = 2\pi\delta\left(\alpha_3+\alpha_4-\alpha_1-\alpha_2-\frac{u+v-\tilde{u}-\tilde{v}}{\omega''}\right),
\eeq
and
\beq
I_{\theta} = 2\pi\delta_{l+s,m+n},
\eeq
respectively. Substituting these results in the expression for $I$ above, we can define the vertex function as $\chi_{\{\vett{m}\}}(P)=(\hbar^2\chi/2)I(P)$, leading to
\barr
\chi_{\{\vett{m}\}}(P)&=&2\pi^2\hbar^2\chi\delta_{l+s,m+n}\int_0^{\infty} dr\, r\int_0^{\infty} d^4\alpha\vett{F}_{pqrt}(\alpha)\nonumber\\
&\times&J_m\left(\xi\alpha_1r\right)J_n\left(\xi\alpha_2r\right)J_l\left(\xi\alpha_3r\right)J_s\left(\xi\alpha_4r\right)\nonumber\\
&\times&\delta\left(\alpha_3+\alpha_4-\alpha_1-\alpha_2-\frac{P}{\omega''}\right),
\earr
where $\vett{F}_{pqrt}(\alpha)=f_p(\alpha_1)f_q(\alpha_2)_r(\alpha_3)f_t(\alpha_4)$. For later convenience, it is instructive to calculate the Fourier transform of the above equation. To to that, we first define the function $\mathcal{K}_{mnls}(r,\alpha,\xi)=J_m\left(\xi\alpha_1r\right)J_n\left(\xi\alpha_2r\right)J_l\left(\xi\alpha_3r\right)J_s\left(\xi\alpha_4r\right)$, so that we can define
\beq
K_{mnls}(\xi,\alpha)=\int_0^{\infty} dr\,r\mathcal{K}_{mnls}(r,\alpha,\xi).
\eeq
Then, the Fourier transform $\sigma_{\{\vett{m}\}}(X)$ of the vertex function $\chi_{\{\vett{m}\}}(P)$ can be written as
\begin{widetext}
\beq
\sigma_{\{\vett{m}\}}(X)=\frac{1}{2\pi\omega'}\int\,dP\chi_{\{\vett{m}\}}(P)e^{iPX/\omega''}=\frac{\pi\hbar^2\chi}{\omega''}\delta_{l+s,m+n}\int_0^{\infty} d^4\alpha K_{mnls}(\xi,\alpha)\vett{F}_{pqrt}(\alpha)e^{i(\alpha_3+\alpha_4-\alpha_1-\alpha_2)X}.
\eeq
\end{widetext}
The $\alpha$-integral in the above expression can be further simplified by noticing that each $\alpha_k$-integral can be solved individually and it just amounts to the definition of generalised OAM-carrying X wave given by Eq. \eqref{eq13}, evaluated at $\{\theta=0 ,\zeta=X\}$ and with $v=0$,namely 
\beq
\int_0^{\infty}d\alpha_kf_{\mu}(\alpha_k)J_{\nu}(\xi\alpha_k r)e^{i\alpha_kX}=\fieldXo{\mu}{\nu}.
\eeq
Using this fact, we can then rewrite the Fourier transform of the vertex function as
\barr
\sigma_{\{\vett{m}\}}(X)&=&\frac{\pi\hbar^2\chi}{\omega''}\delta_{m+n,l+s}\int_0^{\infty}dr\,r\fieldXco{m}{p}\nonumber\\
&\times&\fieldXco{n}{q}\fieldXo{l}{r}\fieldXo{s}{t}.
\earr 
The vertex function can be therefore written in terms of its Fourier transform as follows:
\beq\label{fourierChi}
\chi_{\{\vett{m}\}}(P)=\int d\eta\sigma_{\{\vett{m}\}}(\eta)e^{-i\frac{\eta P}{\omega''}}.
\eeq

\section*{Appendix C: Derivation of Eq. (70)}
By applying the low velocity limit to Eq. \eqref{Hexact}, we have that $\sqrt{\omega_{mp}(v_1)\omega_{nq}(v_2)\omega_{lr}(v_3)\omega_{st}(v_4)}\simeq\omega^2$. Then, after having substituted the creation and annihilation operators $\creatX{m}{p}{v}$ and $\annX{m}{p}{v}$ with their Fourier transforms as prescribed by Eqs. \eqref{opFourier}, the Kerr Hamiltonian has the following form
\barr
\hat{H}_I&=&\frac{\omega^2}{2}\sum_{\multi}\int\,d^4\eta\,\creatF{m}{p}{\eta_1}\creatF{n}{q}{\eta_2}\annF{l}{r}{\eta_3}\annF{s}{t}{\eta_4}\nonumber\\
&\times&\int\,d^4v\chi_{\multi}(\nu)e^{i\vett{v}\cdot\boldsymbol\eta/\omega''},
\earr
where $\vett{v}\cdot\boldsymbol\eta=-v_1\eta_1-v_2\eta_2+v_3\eta_3+v_4\eta_4$ and $\nu=-v_1-v_2+v_3+v_4$. In this appendix, we will calculate explicitly the integral over $d^4v$ and show, that the final expression for the Kerr Hamiltonian is given by Eq. \eqref{eq70}. To do that, we use the expression for the vertex function $\chi_{\multi}(\nu)$ as given by Eq. \eqref{fourierChi} and notice that, by doing so, the above integral in $d^4v$ can be written as the product of four independent integrals in the four variables $v_k$ ($k\in\{1,2,3,4\}$). Moreover, these integral are all of the form
\beq
\int\,dv\,e^{iv\eta/\omega''}=2\pi\omega''\delta(\eta).
\eeq
Using this result in the above equation for each of the $v_k$-integrals leads to the following definition for the quantity $\Sigma_{\multi}(\eta)$:
\begin{widetext}
\beq
\Sigma_{\multi}(\eta)=(2\hbar\omega\omega''\pi^2)^2\chi\int_0^{\infty}dr\,r\fieldXco{m}{p}\fieldXco{n}{q}\fieldXo{l}{r}\fieldXo{s}{t}.
\eeq
\end{widetext}
Direct substitution of this quantity into the above expression for $\hat{H}_I$ gives exactly Eq. \eqref{eq70}.


\begin{thebibliography}{99}
\bibitem{note24} In this context, the bosonic nature of light has been implicitly assumed. However, in general, one could as well define a theory of quantum X waves for fermionic systems. In this case, then, the fermionic, rather than the bosonic, commutation relation must be used to correctly quantise the correspondent harmonic oscillator.
\bibitem{comment1} The velocity of an X wave is related to the characteristic Bessel cone angle $\vartheta_0$ by the relation $v=c/\cos\vartheta_0$.
\bibitem{comment2}Note that in deriving the expression for the expectation value of the Hamilton operator we employed its normal ordered form, thus neglecting the extra $1/2$ factor appearing in Eq. \eqref{ham2}. This to emphasise that even in the normal ordered case, the total energy is still infinite since X waves carry infinite energy.
\bibitem{nota2}As the velocity depends on the Bessel opening angle, moreover, the low velocity approximation can be also viewed as the paraxial approximation.
\end{thebibliography}
\end{document}